\documentclass[aps,pre,twocolumn,superscriptaddress,showpacs,longbibliography]{revtex4-1}
\usepackage{amssymb,amsfonts,amsmath,amsthm}

\usepackage[dvips]{graphicx}
\usepackage{color}

\begin{document}

\title{Comment on ``Ratchet universality in the presence of thermal noise''}

\author{Niurka R.\ Quintero}
\email{niurka@us.es}
\affiliation{Instituto de Matem\'aticas de la Univesidad de Sevilla (IMUS)}
\affiliation{Departamento de F\'\i sica Aplicada I, E.P.S., Universidad de
Sevilla, Virgen de \'Africa 7, 41011, Sevilla, Spain}

\author{Renato Alvarez-Nodarse}
\email{ran@us.es}
\affiliation{Instituto de Matem\'aticas de la Univesidad de Sevilla (IMUS)}
\affiliation{Departamento de An\'alisis Matem\'atico, Universidad de Sevilla,
Apdo 1160, 41080, Sevilla, Spain}

\author{Jos\'e A.\ Cuesta}
\email{cuesta@math.uc3m.es}
\affiliation{Grupo Interdisciplinar de Sistemas Complejos (GISC), Departamento
de Matem\'aticas, Universidad Carlos III de Madrid, Avda.~de la Universidad 30,
28911 Legan\'es, Spain}
\affiliation{Instituto de Biocomputaci\'on y F\'{\i}sica de Sistemas Complejos
(BIFI), Universidad de Zaragoza, 50009 Zaragoza, Spain}

\date{\today}

 
\begin{abstract} 
A recent paper [Phys.~Rev.~E \textbf{87}, 062114 (2013)] presents numerical
simulations on a system exhibiting directed ratchet transport of a driven
overdamped Brownian particle subjected to a spatially periodic, symmetric
potential. The authors claim that their simulations prove the existence of
a universal waveform of the external force which optimally enhances directed
transport, hence confirming the validity of a previous conjecture put forward
by one of them in the limit of vanishing noise intensity. With minor
corrections due to noise, the conjecture holds even in the presence of noise,
according to the authors. On the basis of their results the authors claim
that all previous theories, which predict a different optimal force waveform,
are incorrect. In this comment we provide sufficient numerical evidence showing
that there is no such universal force waveform and that the evidence
obtained by the authors otherwise is due to a fortunate choice of the
parameters. Our simulations also suggest that previous theories correctly
predict the shape of the optimal waveform within their validity regime, namely
when the forcing is weak. On the contrary, the aforementioned conjecture
is shown to be wrong.
\end{abstract}

%
%
\maketitle

The authors of Ref.~\cite{martinez:2013} (see also the
erratum~\cite{martinez:2013b}) simulate the equation
\begin{equation}
\begin{split}
&\dot x+\sin x=\sqrt{\sigma}\xi(t)+\gamma F_{\text{bihar}}(t), \\
&F_{\text{bihar}}(t)=\eta\sin(\omega t)+2(1-\eta)\sin(2\omega t+\phi),
\end{split}
\label{eq:system}
\end{equation}
where $\gamma$ is the global amplitude of the force; $0\leqslant\eta\leqslant
1$ and $\phi$ account for the relative amplitude and initial phase difference
of the two harmonics, respectively; $\xi(t)$ is a Gaussian white noise with
zero mean and $\langle\xi(t)\xi(t + s)\rangle =\delta(s)$; and $\sigma$ is
proportional to the temperature of the system. This system exhibits ratchet
transport if the external force breaks both, a time-shift symmetry, namely if
$F_{\text{bihar}}(t)\ne -F_{\text{bihar}}(t+T/2)$ ($T$ being the period of
$F_{bihar}$), and time-reversal, i.e., $F_{\text{bihar}}(t)\ne
-F_{\text{bihar}}(-t)$. This happens for all $0<\eta<1$ and all $\phi\ne
0,\pi$. If initially the particle starts at $x(t_0)=x_0$, the ratchet current
can be obtained as
\begin{equation}
v=\lim_{t\to\infty}\frac{\langle x(t)\rangle-x_0}{t-t_0},
\label{eq:v}
\end{equation}
where $\langle\cdot\rangle$ represents an ensemble average over all
trajectories satisfying the same initial condition.

Obviously the ratchet current $v$ will be a function of the parameters of the
system, in particular of those that define the external force. Since for
$\eta=0$, $1$ or $\phi=0$, $\pi$ the force neither breaks the time-shift
symmetry nor the time-reversal symmetry (hence $v=0$), it is easily foreseen that
for a certain combination of the parameters of the force $v$ must be maximal
(in absolute value).

Based on a conjecture proposed by one of the authors \cite{chacon:2010}, $v$
should be optimal when the force maximally breaks the symmetries. For $\sigma=0$
this happens for $\eta=4/5$ irrespective of the value of $\gamma$ and of $\phi$
(as long as $\phi\ne 0,\pi$) \cite{martinez:2013}. An argument based on an affine
transformation of the force leads the authors to conclude that this optimal
shape of the force will hold even for $\sigma>0$ ---albeit some deviations are
to be expected.

This result is universal in the sense that is independent of $\gamma$ and
$\phi$. Figure~1(a) of \cite{martinez:2013b} confirms that this is an accurate
prediction even for the high intensities of the noise they use in their
simulations ($\sigma=2$, $3$, $4$). The other parameters are set to
$\omega=0.08\pi$, $\phi=\pi/2$, and $\gamma=2$ throughout their paper.

They go on to claim that, since all previous theories \cite{hanggi:2009,
flach:2000,denisov:2002,wonneberger:1981,marchesoni:1986,quintero:2010} predict
a form of the ratchet current given by \cite{martinez:2013b}
\begin{equation}
v\propto\gamma^3\eta^2(1-\eta),
\end{equation}
they all predict that $v$ is optimal for $\eta=2/3$, a value certainly far
away from the simulation results.

Accordingly the two main conclusions of this work are: (i) the conjecture
of a universal force waveform which optimizes the current is confirmed even
in the presence of strong noise ---albeit with some deviations---, and (ii)
all previous theories must be incorrect because they incorrectly predict this
form.

\begin{figure}
\includegraphics[width=3.2in,clip=]{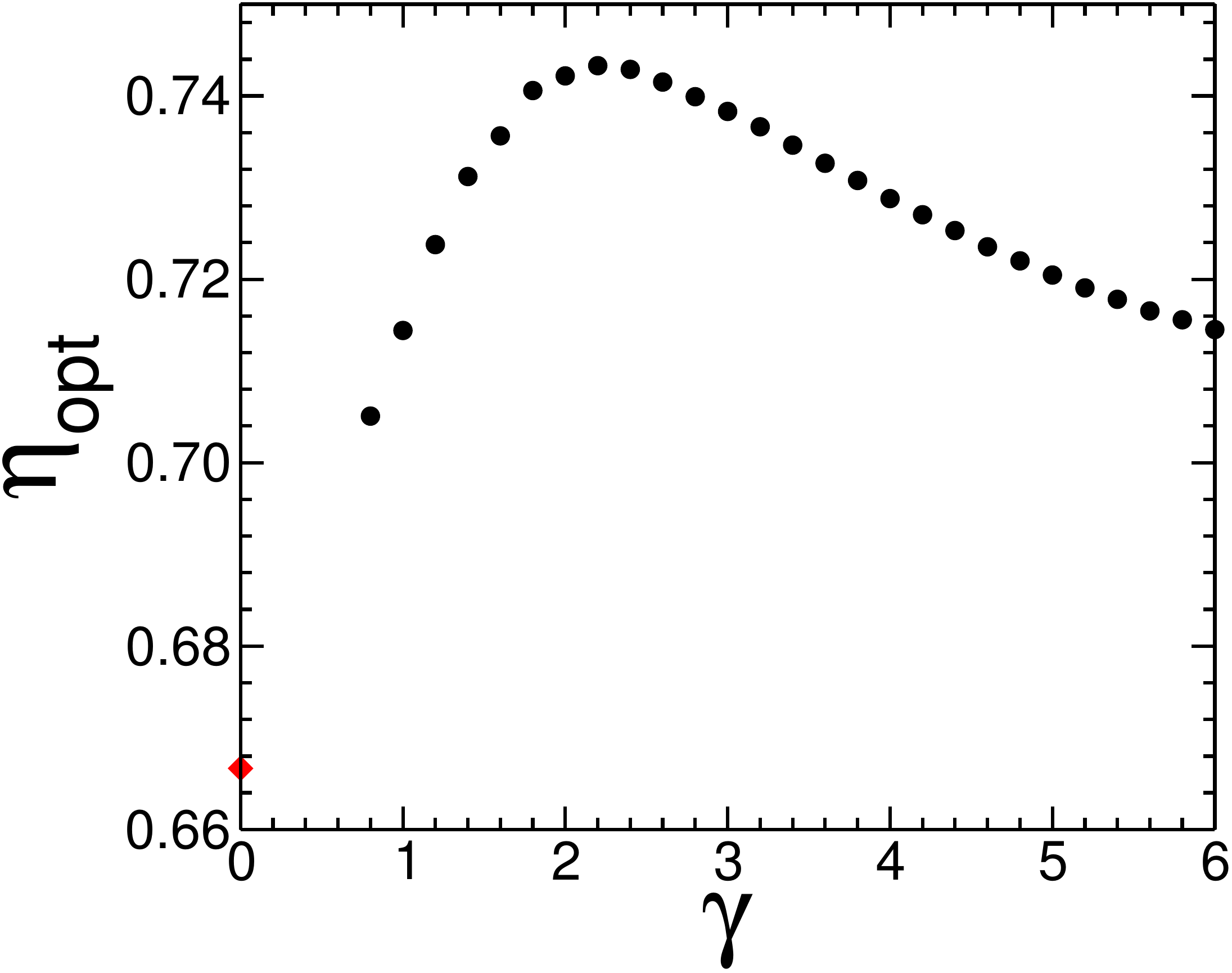}
\caption{(Color online) Values of the parameter $\eta$ defining the relative
amplitudes of the two harmonics of the external force $F_{\text{bihar}}(t)$
[c.f.~Eq.~\eqref{eq:system}], for which the ratchet velocity $v$ reaches its
maximum absolute value, plotted as a function of the global amplitude of the
external force $\gamma$ (full black dots). The remaining parameters are
chosen as in Fig.~1 of \cite{martinez:2013} (actually of the erratum
\cite{martinez:2013b}): $\omega=0.08\pi$, $\sigma=2$, $\phi=\pi/2$. The red
diamond at $(0,2/3)$ represents the theoretical prediction of previous theories
\cite{hanggi:2009,flach:2000,denisov:2002,wonneberger:1981,marchesoni:1986,
quintero:2010}, which should hold in the limit $\gamma\to 0$.}
\label{fig:etagamma}
\end{figure}

We have carried out extensive simulations of the same system \eqref{eq:system}
and with the same parameter as the authors of
\cite{martinez:2013,martinez:2013b}, but instead of limiting ourselves to the
single value of the global amplitude $\gamma=2$ used in their simulations we
have covered a wider range of values, from $\gamma=6$ down to $\gamma=0.8$.
Below this value simulations are prohibitively long because the high values of
the noise intensity demand a very large number of realizations to achieve
reliable results. The outcome of these simulations is summarized in
Fig.~\ref{fig:etagamma}, which represents the value of $\eta$ (henceforth
$\eta_{\text{opt}}$) which optimizes $v$ as a function of the global amplitude
of the external force $\gamma$.

There are three main conclusions that we can extract from this figure. First of
all, there is no such thing as an optimal force waveform. The values of
$\eta_{\text{opt}}$ range from near $0.69$ up to near $0.75$. The predicted
universal value $\eta_{\text{opt}}=4/5$ is reached at no value of $\gamma$,
and the closest it gets to it is at $\gamma\approx 2$ ---precisely the
value used in the simulations of Ref.~\cite{martinez:2013,martinez:2013b}. We
need to make clear at this point that setting $\gamma=2$ in our simulations
our results reproduce accurately the plots of Fig.~1(a) of this work.
This leads us to our second conclusion, namely that the authors of this work
have been misled by their specific choice of the simulation
parameters. Finally, although we cannot decrease $\gamma$ below $0.8$ without
introducing too much uncertainty, the figure clearly illustrates that the trend
of the value of $\eta$ which optimizes $v$ is toward the value $2/3$ which all
theories predict in their range of validity, i.e., \emph{in the limit of weak
external forces.}

On the basis of this evidence we conclude that the conjecture put forward in
\cite{chacon:2010} is wrong, no matter how appealing it may sound. The
reasoning leading from maximum symmetry breaking of the external force to a
maximum response of the system is of a ``linear response'' style, and does not
hold for the kind of nonlinear behavior that ratchet current generation
represents.

For the sake of reproducibility we provide the details of the numerical
procedure we have followed to obtain Fig.~\ref{fig:etagamma}.
Simulations of the stochastic differential equation \eqref{eq:system} have been
performed using the 2nd order weak predictor-corrector method
\cite{kloeden:1995} with time-step $\Delta t=0.01$, initial condition $x(0)=0$,
and a final integration time $t_{f}=200\pi/\omega$. The ratchet velocity $v$ has
been computed using formula \eqref{eq:v} averaging over 5000 realizations of the
noise. For each $\gamma$ in Fig.~\ref{fig:etagamma} we have obtained an entire
curve $v(\eta)$ for $100$ values of $\eta$ in the interval $[0,1]$.

\begin{figure}
\includegraphics[width=3.2in,clip=]{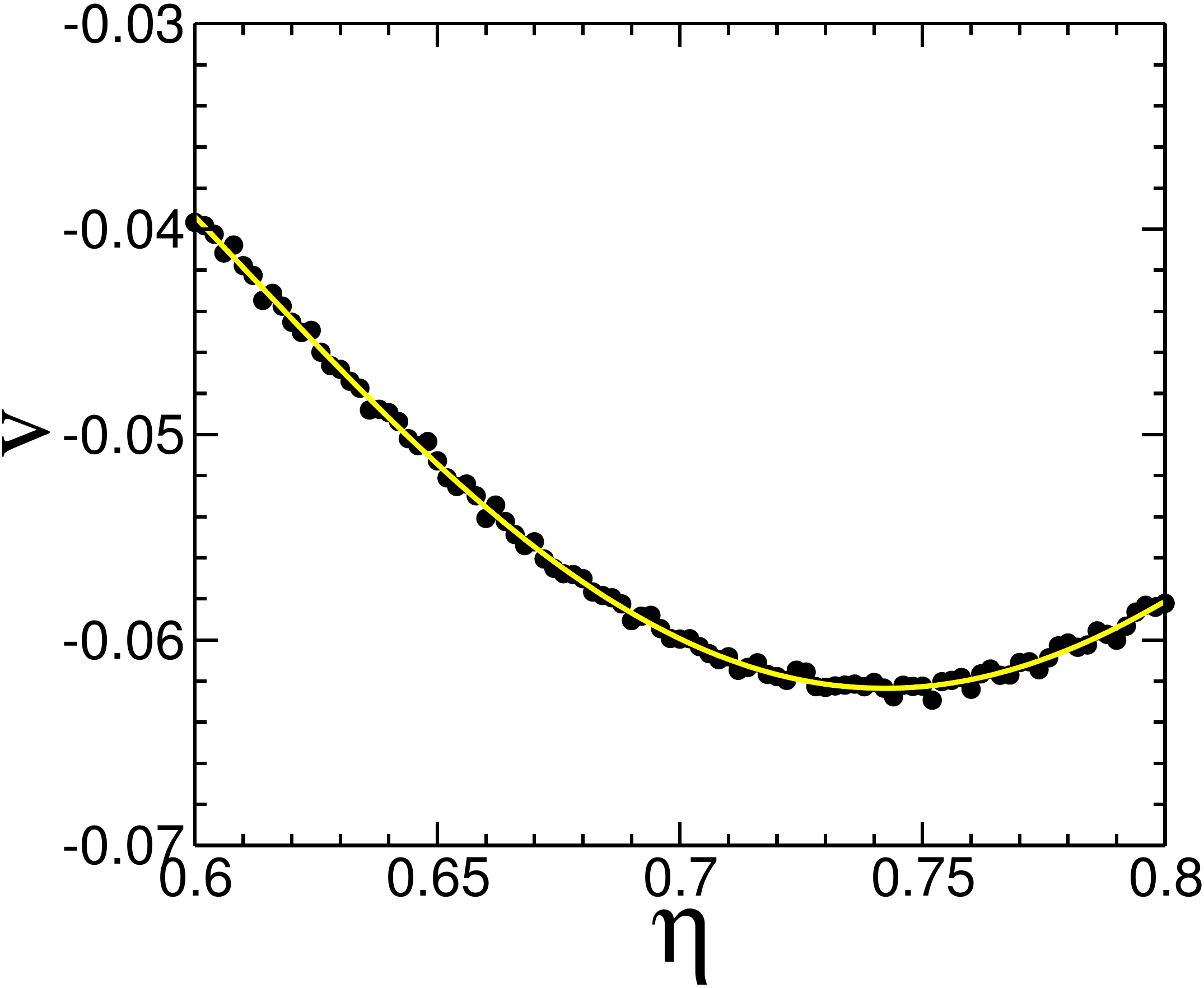}
\caption{(Color online) Value of the ratchet current $v$ as a function of the
parameter $\eta$ defining the relative amplitudes of the two harmonics of the
external force $F_{\text{bihar}}(t)$ [c.f.~Eq.~\eqref{eq:system}]. Parameters:
$\gamma=2$, $\omega=0.08\pi$, $\sigma=2$, $\phi=\pi/2$. Bullets are the
results from simulations averaged over 5000 realizations of the noise. The
yellow curve is a fourth order polynomial fit to these results ($v=-3.7188 + 22.068
\eta - 48.018 \eta^2 + 44.984 \eta^3 - 15.366 \eta^4$; correlation coefficient
$r=0.99$).  This fit is used to determine $\eta_{\text{opt}}$. Notice that the
minimum of the fit (at $\eta=0.742$) does not coincide with the value of $\eta$
for which the largest absolute value of $v$ occurs because of the fluctuations
in the ratchet current.}
\label{fig:veta}
\end{figure}

Despite the average over such a large number of realizations, the resulting
curves are still quite noisy ---too much to reliably determine the value
$\eta_{\text{opt}}$. For this reason we have recalculated the curves $v(\eta)$
for another $100$ values of $\eta$ in a narrower interval that clearly contains
$\eta_{\text{opt}}$, and have fitted a fourth degree polynomial to the results
(see Fig.~\ref{fig:veta} for an example). The value of $\eta_{\text{opt}}$ is
obtained by optimizing this polynomial. This is how the points of
Fig.~\ref{fig:etagamma} have been obtained.

\begin{figure}
\includegraphics[width=3.2in,clip=]{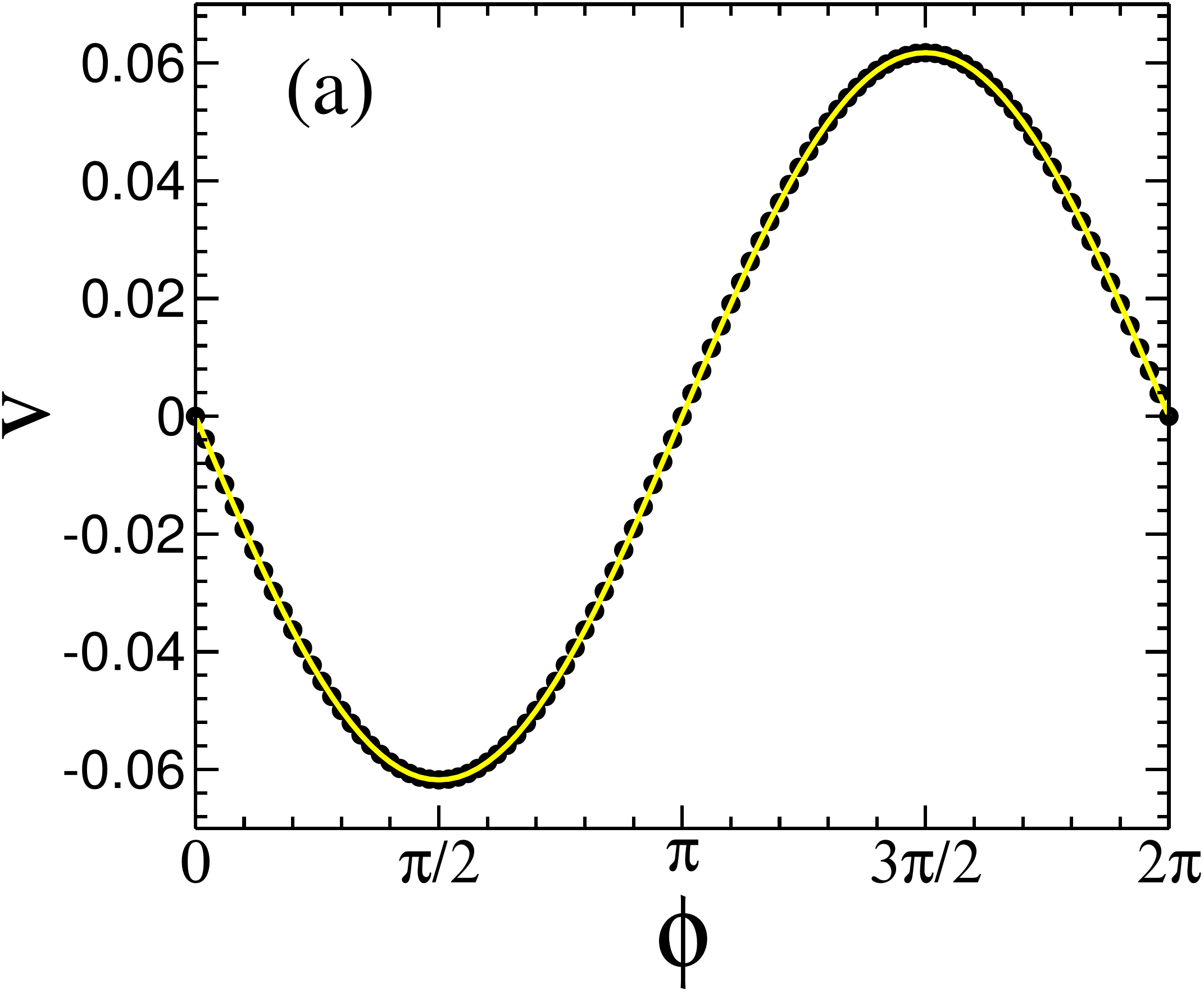}\\
\includegraphics[width=3.2in,clip=]{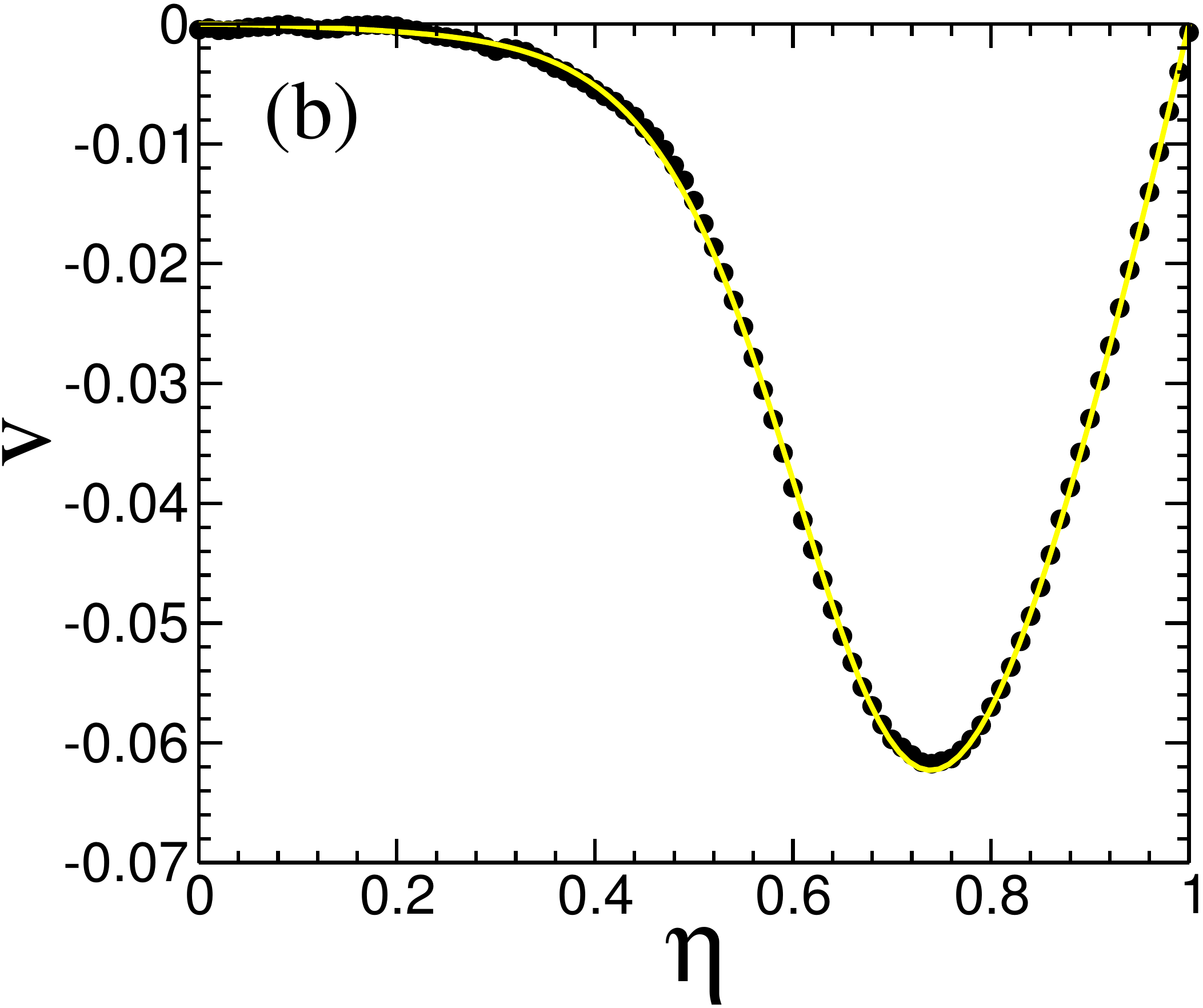}
\caption{(Color online) (a) $v$ as a function of $\phi$ for $\eta=2/3$, (b)
ratchet current $v$ as a function of $\eta$, for $\phi=\pi/2$. Bullets are the
numerical simulations of Eq.~\eqref{eq:system} with parameters
$\omega=0.08\pi$, $\sigma=2$, and $\gamma=2$. The yellow solid line in (a) is a
fit to a sinusoidal function ($v=-0.062 \sin\phi$). The yellow solid line in
(b) is a fit to $-\eta^2(1-\eta)A(\eta)$, where $A(\eta)$ is the $(2,2)$-Pad\'e
approximant
$A(\eta)=(0.025-0.113\eta+0.212\eta^2)(1-2.662\eta+2.011\eta^2)^{-1}$.}
\label{fig:extended}
\end{figure}

As for the second conclusion of Ref.~\cite{martinez:2013}, aside from the
evidence provided by Fig.~\ref{fig:etagamma} that the numerical results are
consistent with the $\eta_{\text{opt}}=2/3$ prediction of the theories in the
limit of weak external forces, we can actually go further and show that a
recent extension of the theory developed in Ref.~\cite{quintero:2010}, valid
for arbitrarily large forces \cite{cuesta:2013}, fits perfectly with the
results presented in \cite{martinez:2013,martinez:2013b}. For the case of
harmonic mixing represented by Eq.~\eqref{eq:system}, the theory predicts that
$v$ is given by the harmonic expansion
\begin{equation}
v=\sum_{n=0}^{\infty}A_n(\gamma,\eta)\gamma^{6n+3}\eta^{4n+2}(1-\eta)^{2n+1}
\sin[(2n+1)\phi],
\label{eq:expansion}
\end{equation}
where the coefficients $A_n(\gamma,\eta)$ are functions of the squares of the
amplitudes of the forcing harmonics, i.e., of $\gamma^2\eta^2$ and
$\gamma^2(1-\eta)^2$. This implies that, if the current is well described by
one sinusoidal function, then
\begin{equation}
v=A_0(\gamma,\eta)\gamma^3\eta^2(1-\eta)\sin\phi+O(\gamma^9),
\label{eq:leading}
\end{equation}
and $A_0(\gamma,\eta)$ should be well described by a bivariate quadratic
polynomial ---or any other approximant of an equivalent order--- in
$\gamma^2\eta^2$ and $\gamma^2(1-\eta)^2$.

Figure~\ref{fig:extended} (top) shows a fit of a sinusoidal function to the
simulation data for $v$ as a function of $\phi$ obtained from
Eq.~\eqref{eq:system} for $\eta=2/3$ and the other parameters as in Fig.~1 of
\cite{martinez:2013b}. It clearly shows that retaining only the first harmonic
in \eqref{eq:expansion} is enough to accurately reproduce the data. Thus $v$
should conform to \eqref{eq:leading}. Accordingly, we set $\phi=\pi/2$ and fit
the simulation results of $v$ vs.~$\eta$ to a function of the form
$-\eta^2(1-\eta)A(\eta)$, where we take for $A(\eta)$ a $(2,2)$-Pad\'e
approximant \footnote{The only reason to use a Pad\'e approximant instead of a
polynomial is that rational approximants are less prone to introduce spurious
oscillations than high degree polynomials. The choice of a $(2,2)$-Pad\'e is
dictated by its having as many unknowns as a fourth degree polynomial ---so
they both are approximants of the same order.}.  The result is plotted in
Fig.~\ref{fig:extended} (bottom) to show that this fit is a very accurate
description of $v(\eta)$ ---and therefore correctly predicts the deviation of
$\eta_{\text{opt}}$ from its weak force approximation $\eta_{\text{opt}}=2/3$.

\begin{figure}
\includegraphics[width=3.2in,clip=]{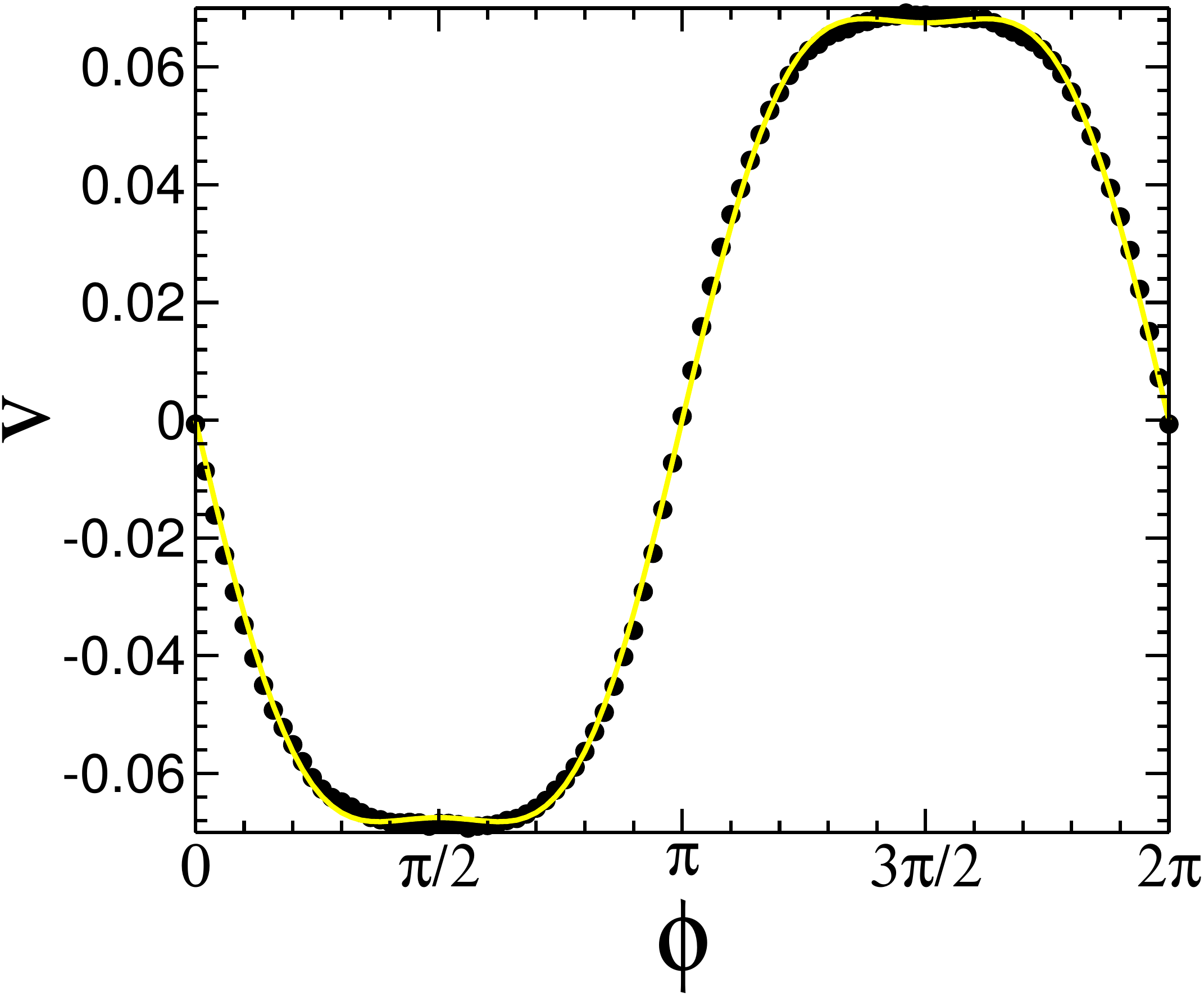}
\caption{(Color online) $v$ as a function of $\phi$ for $\eta=2/3$, $\gamma=6$,
$\omega=0.08\pi$, and $\sigma=2$. Bullets are the numerical simulations of
Eq.~\eqref{eq:system}. The yellow solid line in the top panel is a fit to the
two first harmonics of Eq.~\eqref{eq:expansion} ($v=-0.079 \sin\phi-0.011\sin
3\phi$).}
\label{fig:newharmonics}
\end{figure}

Finally we would like to point out that the idea of an optimal shape of the
external force is very difficult to reconcile with the current shape given by
Eq.~\eqref{eq:expansion}, because as soon as the amplitude of the force
$\gamma$ becomes sufficiently large, new harmonics will start modulating the
shape of the current (Fig.~\ref{fig:newharmonics} clearly illustrates this
effect). In this regime, only a very specific dependence of the coefficients
$A_n$ with $\gamma$ ---which does not occur in the case of
Eq.~\eqref{eq:system}--- would yield the universality claimed in
\cite{chacon:2010,martinez:2013,martinez:2013b}.

\begin{acknowledgments}
We acknowledge financial support through grants MTM2012-36732-C03-03 (R.A.N.),
FIS2011-24540 (N.R.Q.), and PRODIEVO (J.A.C.), from Ministerio de Econom\'{\i}a
y Competitividad (Spain), grants FQM262 (R.A.N.), FQM207 (N.R.Q.), FQM-7276,
and P09-FQM-4643 (N.R.Q., R.A.N.), from Junta de Andaluc\'{\i}a (Spain),
project MODELICO-CM (J.A.C.), from Comunidad de Madrid (Spain), and a grant
from the Humboldt Foundation through Research Fellowship for Experienced
Researchers SPA 1146358 STP (N.R.Q.).
\end{acknowledgments}

\bibstyle{apsrev4-1}
\bibliography{ratchets}

\end{document}